\documentclass[sigconf]{acmart}
\usepackage{xcolor}
\usepackage{subfigure}

\definecolor{mycolor}{RGB}{127,185,87}

\usepackage{colortbl} % 提供 \columncolor 命令
\usepackage{xcolor}   % 提供颜色定义

\usepackage{multirow}
\usepackage[normalem]{ulem}
\useunder{\uline}{\ul}{}
\usepackage{enumitem}
\usepackage{epigraph} % 引言排版包
\AtBeginDocument{%
  }

\setcopyright{acmlicensed}
\copyrightyear{2025}
\acmYear{2025}
\acmDOI{XXXXXXX.XXXXXXX}
\acmConference[MM '25]{Proceedings of the 33rd ACM International Conference on Multimedia}{October 27--31, 2025}{Dublin, Ireland}
\acmISBN{978-1-4503-XXXX-X/2018/06}

\acmSubmissionID{xxx}

\begin{document}

%%
%% The "title" command has an optional parameter,
%% allowing the author to define a "short title" to be used in page headers.
\title{MER 2025: When Affective Computing Meets Large Language Models}

\author{Zheng Lian}
\affiliation{%
	\institution{Institute of Automation, Chinese Academy of Sciences (CAS)}
	\city{Beijing}
	\country{China}
}

\author{Rui Liu}
\affiliation{%
	\institution{Inner Mongolia University}
	\city{Hohhot}
	\country{China}
}

\author{Kele Xu}
\affiliation{%
	\institution{National University of Defense Technology}
	\city{Changsha}
	\country{China}
}

\author{Bin Liu}
\affiliation{%
	\institution{Institute of Automation, CAS}
	\city{Beijing}
	\country{China}
}

\author{Xuefei Liu}
\affiliation{%
	\institution{Tianjin Normal University}
	\city{Tianjin}
	\country{China}
}

\author{Yazhou Zhang}
\affiliation{%
	\institution{Tianjin University}
	\city{Tianjin}
	\country{China}
}

\author{Xin Liu}
\affiliation{%
	\institution{Lappeenranta-Lahti University of Technology}
	\city{Lappeenranta}
	\country{Finland}
}

\author{Yong Li}
\affiliation{%
	\institution{Southeast University}
	\city{Nanjing}
	\country{China}
}

\author{Zebang Cheng}
\affiliation{%
	\institution{Shenzhen University}
	\city{Shenzhen}
	\country{China}
}

\author{Haolin Zuo}
\affiliation{%
	\institution{Inner Mongolia University}
	\city{Hohhot}
	\country{China}
}

\author{Ziyang Ma}
\affiliation{%
	\institution{Shanghai Jiaotong University}
	\city{Shanghai}
	\country{China}
}

\author{Xiaojiang Peng}
\affiliation{%
	\institution{Shenzhen Technology University}
	\city{Shenzhen}
	\country{China}
}

\author{Xie Chen}
\affiliation{%
	\institution{Shanghai Jiaotong University}
	\city{Shanghai}
	\country{China}
}

\author{Ya Li}
\affiliation{%
	\institution{Beijing University of Posts and Telecommunications}
	\city{Beijing}
	\country{China}
}

\author{Erik Cambria}
\affiliation{%
	\institution{Nanyang Technological University}
	\country{Singapore}
}

\author{Guoying Zhao}
\affiliation{%
	\institution{University of Oulu}
	\city{Oulu}
	\country{Finland}
}

\author{Björn W. Schuller}
\affiliation{%
	\institution{Technical University of Munich}
	\city{Munich}
	\country{Germany}
}

\author{Jianhua Tao}
\affiliation{%
	\institution{Tsinghua University}
	\city{Beijing}
	\country{China}
}

%%
%% By default, the full list of authors will be used in the page
%% headers. Often, this list is too long, and will overlap
%% other information printed in the page headers. This command allows
%% the author to define a more concise list
%% of authors' names for this purpose.
\renewcommand{\shortauthors}{Trovato et al.}

%%
%% The abstract is a short summary of the work to be presented in the
%% article.
\begin{abstract}

   MER2025 is the third year of our MER series of challenges, aiming to bring together researchers in the affective computing community to explore emerging trends and future directions in the field. Previously, MER2023\footnote{\emph{http://merchallenge.cn/mer2023}} focused on multi-label learning, noise robustness, and semi-supervised learning, while MER2024\footnote{\emph{https://zeroqiaoba.github.io/MER2024-website}} introduced a new track dedicated to open-vocabulary emotion recognition. This year, MER2025 centers on the theme ``When Affective Computing Meets Large Language Models (LLMs)''. We aim to shift the paradigm from traditional categorical frameworks reliant on predefined emotion taxonomies to LLM-driven generative methods, offering innovative solutions for more accurate and reliable emotion understanding. The challenge contains four tracks: \textbf{MER-SEMI} focuses on fixed categorical emotion recognition enhanced by semi-supervised learning; \textbf{MER-FG} explores fine-grained emotions, expanding recognition from basic to nuanced emotional states; \textbf{MER-DES} incorporates multimodal cues (beyond emotion words) into predictions to enhance model interpretability; \textbf{MER-PR} reveals whether emotion prediction results can improve personality recognition performance. For the first three tracks, the baseline code is available at MERTools\footnote{\emph{https://github.com/zeroQiaoba/MERTools}} and datasets can be accessed via Hugging Face\footnote{\emph{https://huggingface.co/datasets/MERChallenge/MER2025}}. For the last track, the dataset and baseline code are available on GitHub\footnote{\emph{https://github.com/cai-cong/MER25\_personality}}.
   
\end{abstract}


%%
%% Keywords. The author(s) should pick words that accurately describe
%% the work being presented. Separate the keywords with commas.
% \keywords{Do, Not, Us, This, Code, Put, the, Correct, Terms, for,
  % Your, Paper}
% %% A "teaser" image appears between the author and affiliation
% %% information and the body of the document, and typically spans the
% %% page.

% \received{20 February 2007}
% \received[revised]{12 March 2009}
% \received[accepted]{5 June 2009}

%%
%% This command processes the author and affiliation and title
%% information and builds the first part of the formatted document.
\maketitle

\section{Introduction}
Emotion plays a pivotal role in human-computer interactions \cite{picard2000affective, zhou2018emotional}, drawing increasing attention within the field of artificial intelligence. Already today, humans interact more frequently with AI systems than through direct human-to-human interaction, and this trend is expected to grow even more pronounced in the coming years. Emotion-aware interaction systems hold significant potential across various domains, including healthcare, banking, transportation, and education \cite{feidakis2016review,haddad2024emotion}.

Human emotion expression is a complex process that typically involves multiple modalities, including facial expressions, vocal tones, body movements, gestures, and even physiological signals \cite{izard2009emotion}. This complexity has spurred the development of Multimodal Emotion Recognition (MER), a critical task that aims to integrate cross-modal cues to identify human emotions. Recently, MER research has evolved in two key directions: a) from coarse-grained \cite{lian2024merbench} to fine-grained emotion recognition \cite{lian2024open}, and b) from categorical approaches \cite{lian2021ctnet} to descriptive emotion understanding \cite{lian2025affectgpt, cheng2024emotion}, aiming to enhance both prediction accuracy and interpretability. In line with these advancements, MER2025@ACM Multimedia introduces four tracks aligned with current research priorities:

\textbf{Track 1. MER-SEMI.}
Recent studies have demonstrated that pre-training on large-scale unlabeled data \cite{ma2023emotion2vec}, especially domain-matched data, can significantly enhance model performance \cite{lian2024merbench, sun2023mae, cheng2023semi}. This track provides a substantial collection of unlabeled samples from the same domain as the labeled data. Participants are encouraged to leverage semi-supervised learning techniques, such as masked auto-encoders \cite{tongvideomae} or contrastive learning \cite{radford2021learning}, to achieve better results.

\textbf{Track 2. MER-FG.}
Current frameworks primarily focus on basic emotions, often failing to capture the complexity and subtlety of human emotions. This track shifts the emphasis to fine-grained MER, enabling the prediction of a broader range of emotions. Following previous works \cite{lian2024open,lian2025affectgpt}, participants are encouraged to leverage large language models (LLMs) for this purpose. Given that LLMs possess extensive vocabularies, they hold the potential to generate more diverse emotion categories beyond basic labels.

\textbf{Track 3. MER-DES.}
The first two tracks primarily focus on emotion words, neglecting the integration of multimodal clues during the inference process. This omission results in prediction outcomes that lack interpretability. Moreover, emotion words struggle to fully capture the dynamic, diverse, and sometimes ambiguous nature of human emotions. This track seeks to leverage free-form, natural language descriptions to represent emotions \cite{lian2023explainable,lian2025affectgpt}, offering greater flexibility to achieve more accurate emotion representations and enhance model interpretability.

\textbf{Track 4. MER-PR.}
Personality and emotion are deeply intertwined in human behavior and social interactions, yet current research often treats them as separate tasks, neglecting their inherent correlations. This track seeks to investigate the interplay between emotion and personality, exploring whether emotion recognition can enhance the accuracy of personality predictions. Participants are encouraged to employ techniques such as multi-task learning to analyze the influence of emotion on personality prediction.

\begin{figure}[t]
	\centering
	\includegraphics[width=\linewidth]{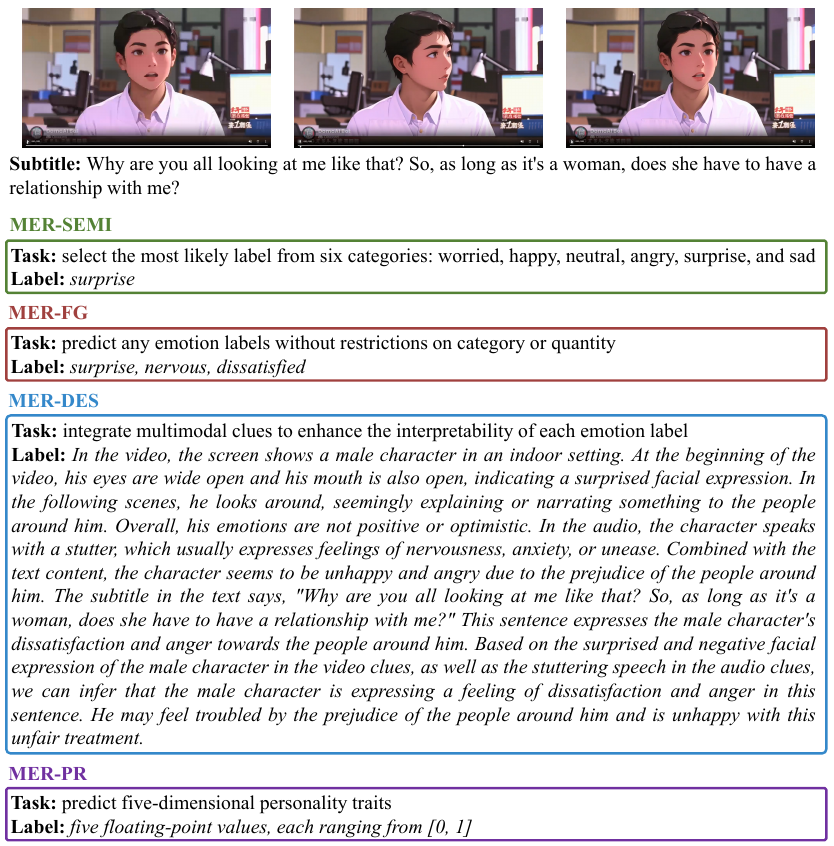}
	\caption{Demo for four tracks.}
	\label{Figure1}
\end{figure}

Figure \ref{Figure1} shows the differences between these tracks. MER2025 builds upon the success of MER 2023@ACM Multimedia \cite{lian2023mer} and MER 2024@IJCAI \cite{lian2024mer}. Over the course of these challenges, participation has steadily grown, increasing from 76 teams in MER2023 to 94 teams in MER2024. This year, we aim to attract more teams to join our challenge. Through the MER series of challenges, we strive to establish a unified platform for comparing different systems and further advancing the development of MER.

\section{MER-SEMI}

\subsection{Dataset}

MER-SEMI spans three consecutive MER challenges, aiming to enhance the performance of categorical emotion recognition algorithms through semi-supervised learning and unlabeled data. This year, we expanded the dataset by incorporating more labeled and unlabeled samples. Our raw data comes from two sources: a) conversational emotion datasets MC-EIU \cite{liu2024emotion} and M3ED \cite{zhao2022m3ed}, with explicit approval from dataset owners; and b) 19 Chinese TV series sourced from publicly available platforms. Then, we follow the video segmentation and filtering process used in prior MER challenges \cite{lian2023mer, lian2024mer}, ensuring each video contains predominantly one speaker with relatively complete speech content. Compared to MER2024 (120k samples), we increased the dataset size to 132k samples in MER2025, introducing richer topic diversity and more characters. Detailed dataset statistics are provided in Table \ref{Table1}.

For evaluation purposes, manually annotating 124k unlabeled data is impractical due to the substantial time and financial costs. Thus, we select a subset of the unlabeled data for performance assessments. To ensure label reliability, we recruit nine annotators and conduct a preliminary test to evaluate their alignment with emotion experts. In this process, two annotators are excluded based on insufficient agreement, leaving seven annotators for the labeling task. Then, we randomly sample 10k instances from the unlabeled subset and ask the annotators to choose the most likely label from eight categories: \emph{neutral}, \emph{happy}, \emph{angry}, \emph{sad}, \emph{surprise}, \emph{worry}, \emph{others}, \emph{unknown}. In this process, each annotator is assigned a portion of the 10k samples to reduce annotation time. Meanwhile, we ensure that each sample is labeled by at least five annotators. Samples that receive at least 80\% agreements and whose majority vote is neither \emph{others} nor \emph{unknown} are included in the test set. This process yields 2,026 high-quality test samples, ensuring strong inter-annotator agreement and reliable ranking results.

\subsection{Evaluation Metrics}
This track aims to identify the most likely label from six candidate emotions: \emph{neutral}, \emph{happy}, \emph{angry}, \emph{sad}, \emph{surprise}, and \emph{worry}. For evaluation purposes, we use the same metrics as previous MER challenges \cite{lian2023mer, lian2024mer}: accuracy and weighted average F1-score (WAF). Given the inherent class imbalance in the dataset, we prioritize WAF as the primary metric for final ranking.

\begin{table}[t]
	\centering
	\caption{Dataset Statistics for MER-SEMI. In MER2025, we further expand the dataset, comprising 7,369 labeled samples and 124,802 unlabeled samples. Additionally, we select 2,026 samples from the unlabeled subset for evaluation.}
	\label{Table1}
        \scalebox{0.9}{
	\begin{tabular}{l|cc|c}
		\hline
		   & Train\&Val & {\begin{tabular}[c]{@{}c@{}}\# of test samples \\ (labeled/whole)\end{tabular}} & Total \\
		\hline 
            MER-SEMI (2023) \cite{lian2023mer} & 3,373 & 834/73,982    & 77,355 \\ 
            MER-SEMI (2024) \cite{lian2024mer}  & 5,030 & 1,169/115,595 & 120,625 \\ 
		MER-SEMI (2025) & 7,369 & 2,026/124,802 & 132,171 \\ 
		\hline
	\end{tabular}
    }
\end{table}

\begin{table}[t]
	\centering
	\renewcommand\tabcolsep{2.4pt}
	\caption{MER-SEMI baseline results (\%). We also report five-fold cross-validation results on the Train$\&$Val set. For multimodal results, Top1 means selecting the best-performing unimodal feature, while Top2 means selecting the two best-performing unimodal features.}
	\label{Table2}
    \scalebox{0.9}{
	\begin{tabular}{l|cc|>{\columncolor[gray]{0.9}}cc}
		\hline
		\multirow{2}{*}{Feature} & \multicolumn{2}{c|}{Train$\&$Val} & \multicolumn{2}{c}{MER-SEMI} \\
		& WAF $(\uparrow)$ & ACC $(\uparrow)$ & WAF $(\uparrow)$ & ACC $(\uparrow)$ \\
		
		\hline
		\multicolumn{5}{c}{Visual Modality} \\
		\hline
ResNet-FER2013 \cite{he2016deep} &57.83$\pm$0.20 & 58.71$\pm$0.34&52.16$\pm$0.25 & 52.90$\pm$0.22 \\
DINOv2-large \cite{oquab2023dinov2} &58.75$\pm$0.11 & 59.77$\pm$0.11&52.43$\pm$0.34 & 53.19$\pm$0.27 \\
SENet-FER2013 \cite{hu2018squeeze}  &57.77$\pm$0.29 & 58.76$\pm$0.22&53.69$\pm$0.08 & 54.90$\pm$0.24 \\
MANet-RAFDB \cite{zhao2021learning} &59.90$\pm$0.15 & 60.87$\pm$0.08&54.31$\pm$0.19 & 54.95$\pm$0.14 \\
CLIP-base \cite{radford2021learning} &61.72$\pm$0.19 & 62.40$\pm$0.22&56.30$\pm$0.19 & 57.53$\pm$0.26 \\
CLIP-large \cite{radford2021learning}&66.58$\pm$0.15 & 66.95$\pm$0.13&60.50$\pm$0.19 & 61.08$\pm$0.15 \\

		\hline
		\multicolumn{5}{c}{Acoustic Modality} \\
		\hline

        WavLM-base \cite{chen2022wavlm}&58.62$\pm$0.11 & 58.91$\pm$0.09&54.55$\pm$0.28 & 55.81$\pm$0.19 \\
wav2vec 2.0-large \cite{baevski2020wav2vec} &67.64$\pm$0.23 & 67.63$\pm$0.20&61.66$\pm$0.27 & 62.76$\pm$0.26 \\
wav2vec 2.0-base \cite{baevski2020wav2vec} &67.55$\pm$0.27 & 67.52$\pm$0.29&62.59$\pm$0.27 & 63.43$\pm$0.30 \\
Whisper-large \cite{radford2023robust} &66.51$\pm$0.15 & 66.56$\pm$0.17&66.52$\pm$0.26 & 67.05$\pm$0.29 \\
HUBERT-base \cite{hsu2021hubert} &72.36$\pm$0.09 & 72.41$\pm$0.08&68.13$\pm$0.26 & 69.05$\pm$0.20 \\
HUBERT-large \cite{hsu2021hubert} &76.29$\pm$0.07 & 76.36$\pm$0.07&72.27$\pm$0.28 & 72.90$\pm$0.36 \\

		\hline
		\multicolumn{5}{c}{Lexical Modality} \\
		\hline

        MacBERT-base  \cite{cui2020revisiting} &53.23$\pm$0.12 & 53.40$\pm$0.15&52.54$\pm$0.13 & 52.56$\pm$0.26 \\
MacBERT-large \cite{cui2020revisiting} &53.68$\pm$0.13 & 53.83$\pm$0.13&52.75$\pm$0.14 & 52.91$\pm$0.29 \\
BLOOM-7B \cite{workshop2022bloom} &54.03$\pm$0.10 & 54.13$\pm$0.15&53.32$\pm$0.32 & 53.08$\pm$0.30 \\
RoBERTa-large \cite{liu2019roberta}&53.80$\pm$0.11 & 54.01$\pm$0.09&53.66$\pm$0.45 & 53.55$\pm$0.42 \\
RoBERTa-base \cite{liu2019roberta} &53.05$\pm$0.08 & 53.30$\pm$0.10&53.78$\pm$0.23 & 53.71$\pm$0.21 \\

            \hline
		\multicolumn{5}{c}{Acoustic + Visual} \\
		\hline
            Top1&80.65$\pm$0.09 & 80.67$\pm$0.09&77.10$\pm$0.44 & 77.54$\pm$0.50 \\
            Top2&80.63$\pm$0.10 & 80.67$\pm$0.13&75.89$\pm$0.23 & 76.28$\pm$0.25 \\

            \hline
		\multicolumn{5}{c}{Acoustic + Lexical} \\
		\hline
            Top1&76.95$\pm$0.22 & 77.06$\pm$0.17&73.64$\pm$0.24 & 74.31$\pm$0.16 \\
            Top2&76.80$\pm$0.07 & 76.90$\pm$0.08&73.57$\pm$0.17 & 74.13$\pm$0.22 \\

            \hline
		\multicolumn{5}{c}{Visual + Lexical} \\
		\hline
            Top1&73.52$\pm$0.18 & 73.56$\pm$0.16&72.09$\pm$0.21 & 72.55$\pm$0.32 \\
            Top2&73.89$\pm$0.12 & 73.98$\pm$0.12&72.13$\pm$0.30 & 72.75$\pm$0.27 \\

            \hline
		\multicolumn{5}{c}{Acoustic + Visual + Lexical} \\
		\hline
            Top1&82.05$\pm$0.11 & 82.10$\pm$0.12&\textbf{78.63}$\pm$0.53 & \textbf{78.77}$\pm$0.55 \\
            Top2&81.82$\pm$0.08 & 81.86$\pm$0.06&77.47$\pm$0.26 & 77.58$\pm$0.25 \\

		\hline

	\end{tabular}
    }
\end{table}

\subsection{Baseline Framework}
A categorical model primarily relies on two key components: feature selection and model architecture. For feature selection, we evaluate the performance of both handcrafted and model-driven features. For model architecture, MERBench highlights that a simple attention mechanism, which maps different unimodal features to the same dimension and then fuses them using attention weights, can already achieve strong performance \cite{lian2024merbench}. In contrast, more complex fusion architectures may lead to overfitting problems and are not suitable for MER, where labeled data is usually limited. For implementation details, please refer to MERBench and our baseline code.

\subsection{Baseline Results}
Our baseline code is designed to automatically and randomly select hyperparameters. In practice, we execute each command 50 times, identify the optimal hyperparameter combination, and then run the code six times under this configuration to report the average results along with the standard deviation. Table \ref{Table2} summarizes the baseline performance for MER-SEMI. To ensure reproducibility, we have included all feature extraction code and pretrained weights in the baseline code. From Table \ref{Table2}, a strong correlation emerges between the five-fold cross-validation results on the Train$\&$Val set and the test set performance. This suggests that participants can use Train$\&$Val results as a reliable indicator for model selection. Notably, trimodal fusion achieves the highest performance, underscoring the complementary value of each modality in emotion recognition. However, for multimodal results, Top2 does not yield better performance than Top1. This is because the fusion process may inadvertently introduce emotion-irrelevant features, potentially degrading overall performance.

\section{MER-FG}

\subsection{Dataset}

\begin{figure}[t]
	\begin{center}
		\subfigure[OV-MERD]{
			\label{Figure2-1}
			\centering
			\includegraphics[width=0.476\linewidth, trim=0 0 0 0]{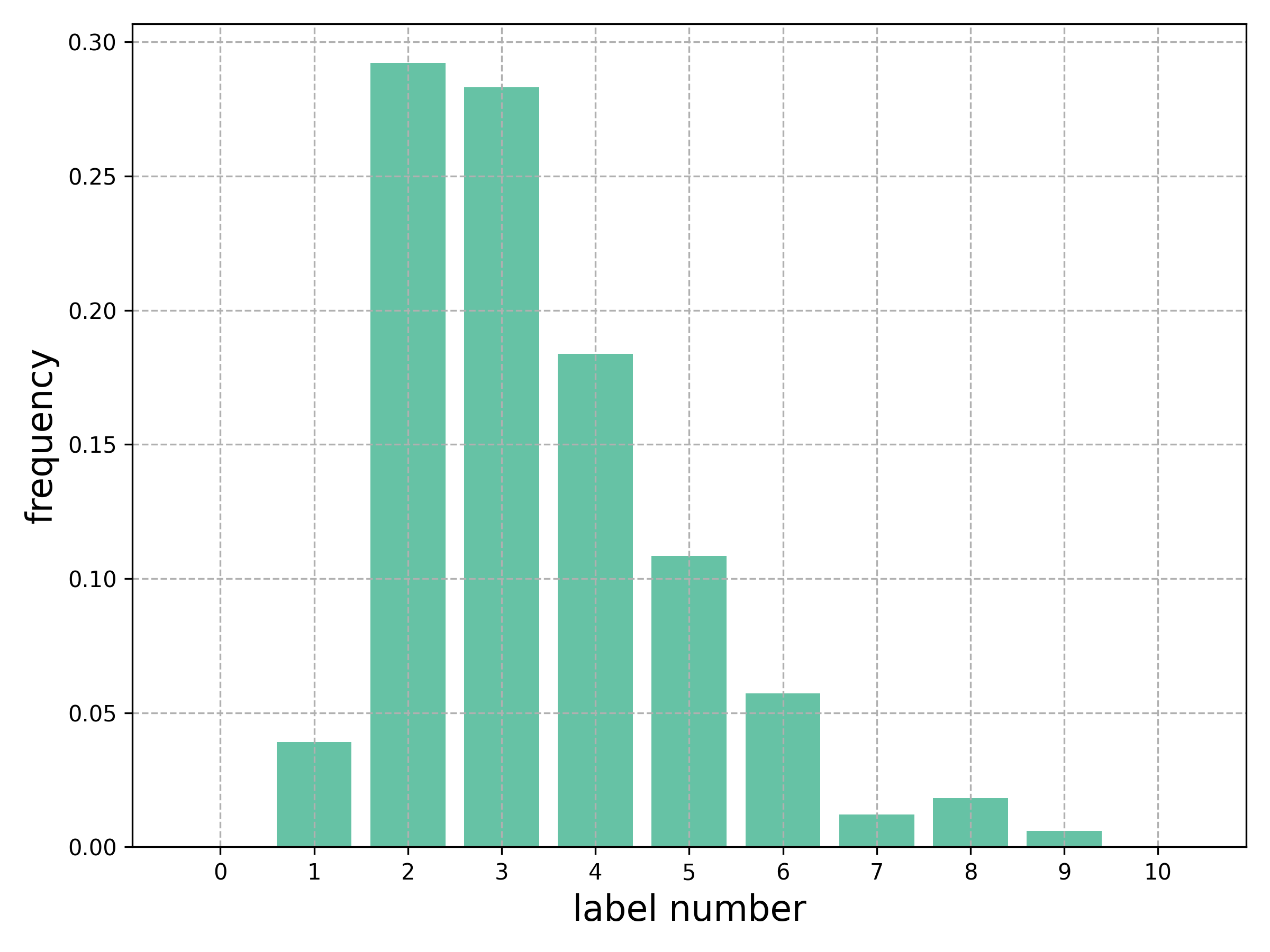}
		}
		\subfigure[MER-Caption+]{
			\label{Figure2-2}
			\centering
			\includegraphics[width=0.476\linewidth, trim=0 0 0 0]{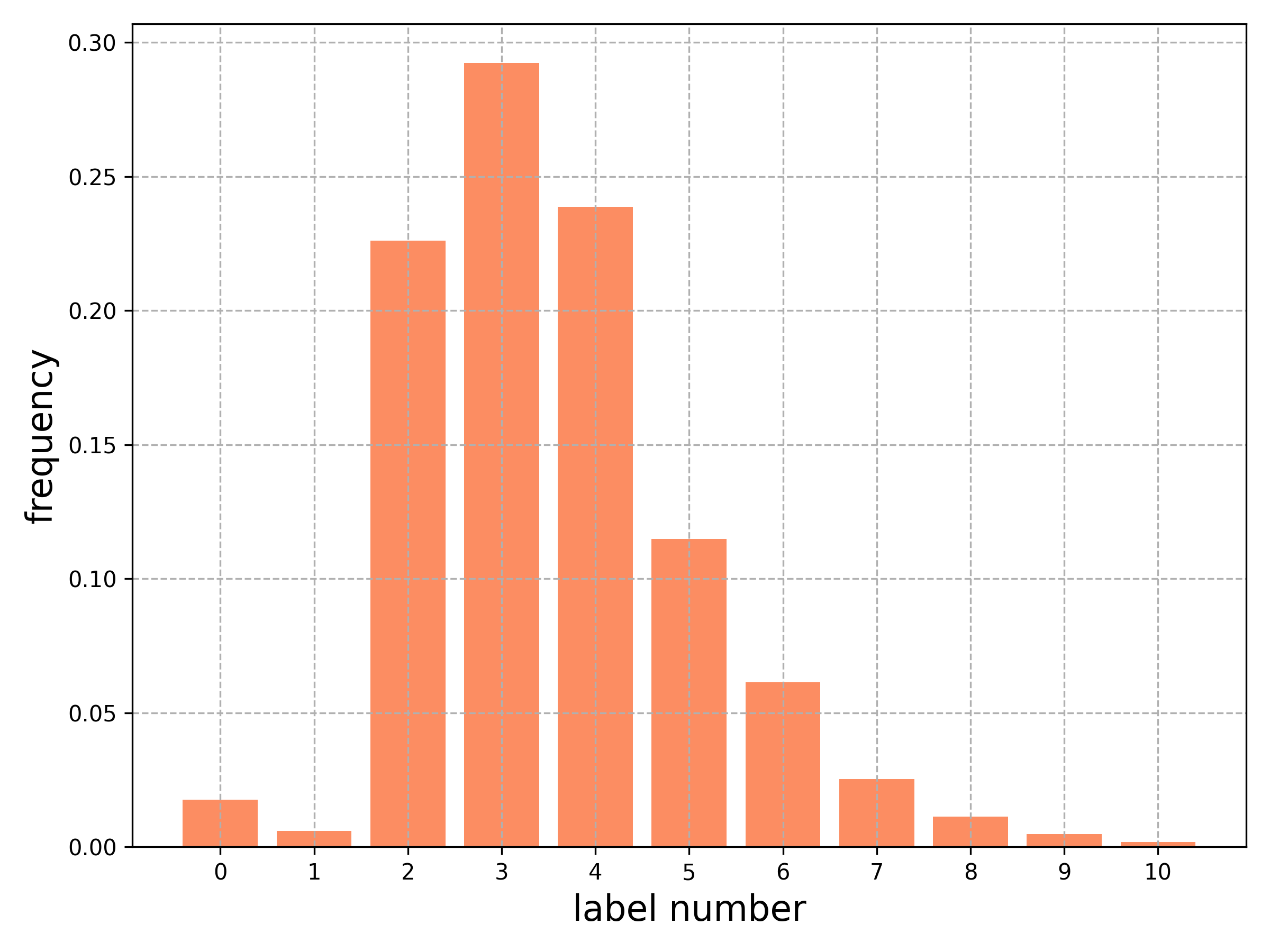}
		}
	\end{center}
	\caption{Label number distribution for MER-FG.}
	\label{Figure2}
\end{figure}

Unlike MER-SEMI, which restricts the prediction scope to six candidates, MER-FG does not limit the label space, allowing predictions for any number and any emotion categories for each sample. In this track, we utilize two recently released datasets, OV-MERD \cite{lian2024open} and MER-Caption+ \cite{lian2025affectgpt}, as training datasets. Their statistics are summarized in Table \ref{Table3}. Specifically, OV-MERD employs a \emph{human-led, model-assisted} annotation strategy, where MLLMs first provide pre-extracted multimodal clues and then rely heavily on manual verification to ensure label quality. In contrast, MER-Caption+ adopts a \emph{human-assisted, model-led} annotation strategy, leveraging human priors to guide description generation and sample filtering, ultimately achieving an automatic annotation process. Consequently, OV-MERD provides small-scale but high-quality labels, whereas MER-Caption+ offers large-scale labels that may contain some errors in emotion annotations. Figure \ref{Figure2} summarizes the sample-wise label number distribution of these datasets. We observe that they provide rich emotion labels for each sample, offering potential for complex emotion modeling.

\begin{table}[t]
	\centering
	\caption{Dataset statistics for MER-FG.}
	\label{Table3}
        \scalebox{0.9}{
	\begin{tabular}{l|c|c}
		\hline
		   & Train\&Val & {\begin{tabular}[c]{@{}c@{}}\# of test samples \\ (labeled/whole)\end{tabular}} \\
		\hline 
            OV-MERD \cite{lian2023mer} & 332 & \multirow{2}{*}{1,200/124,802}\\ 
            MER-Caption+ \cite{lian2024mer}  & 31,327 & \\ 
		\hline
	\end{tabular}
    }
\end{table}

\begin{figure*}[t]
	\centering
	\includegraphics[width=\linewidth]{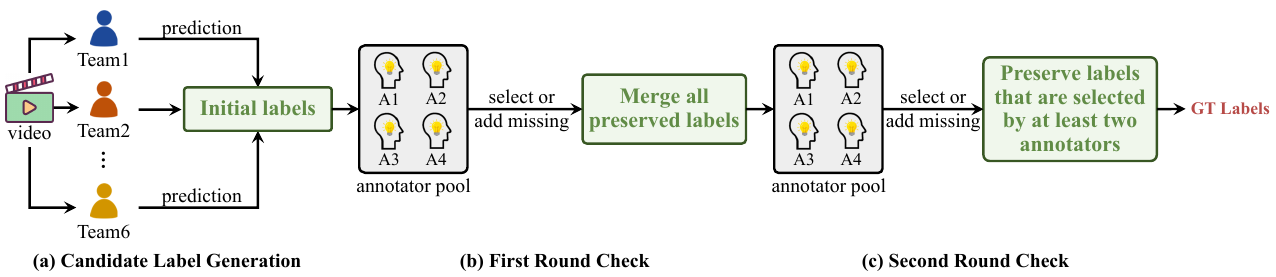}
	\caption{Annotation pipeline for MER-FG.}
	\label{Figure3}
\end{figure*}

For evaluation purposes, we randomly selected 1,200 samples from the unlabeled set in MER-SEMI and engaged four annotators to label them. These annotators were distinct from those involved in the MER-SEMI dataset and had also successfully passed the preliminary qualification test described in MER-SEMI. Figure \ref{Figure3} presents the detailed annotation pipeline. Specifically, we first aggregated the Top-6 teams' submission results from the previous year’s MER-OV track to generate initial labels. Each annotator was then tasked with either selecting labels they deemed correct or adding any additional labels they considered appropriate but not included in the provided candidate list. This approach enabled us to obtain richer emotion annotations for each sample compared to directly asking annotators to provide labels without candidate options. Our manual verification process consisted of two rounds. In the first round, we preserved all labels selected by different annotators to ensure comprehensiveness. In the second round, we retained only those labels confirmed by at least two annotators to ensure accuracy. Through this two-step verification process, each final label was validated at least three times, ensuring both comprehensiveness and accuracy of the annotation results.

\begin{figure}[t]
	\begin{center}
		\subfigure[W1]{
			\label{Figure4-1}
			\centering
			\includegraphics[width=0.42\linewidth, trim=0 0 0 0]{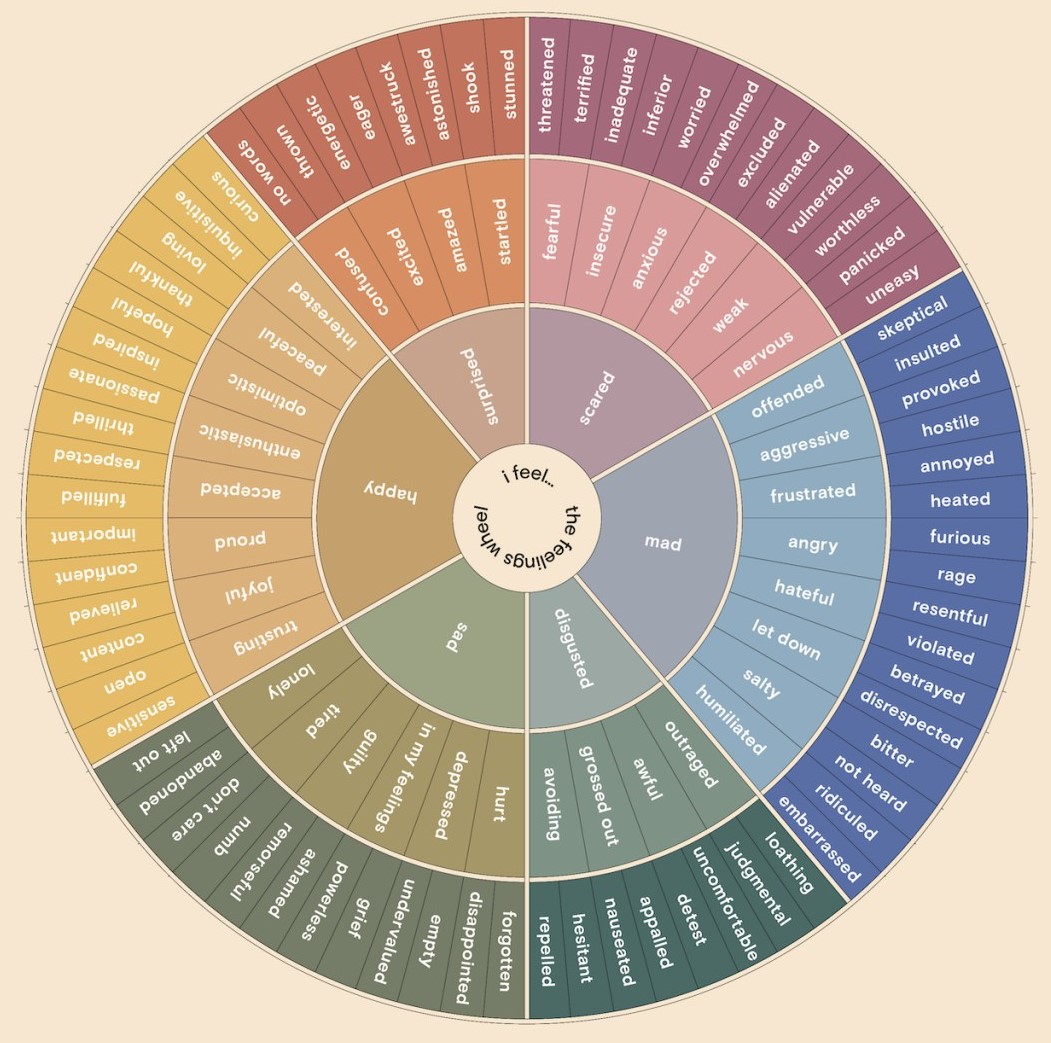}
		}
		\subfigure[W2]{
			\label{Figure4-2}
			\centering
			\includegraphics[width=0.42\linewidth, trim=0 0 0 0]{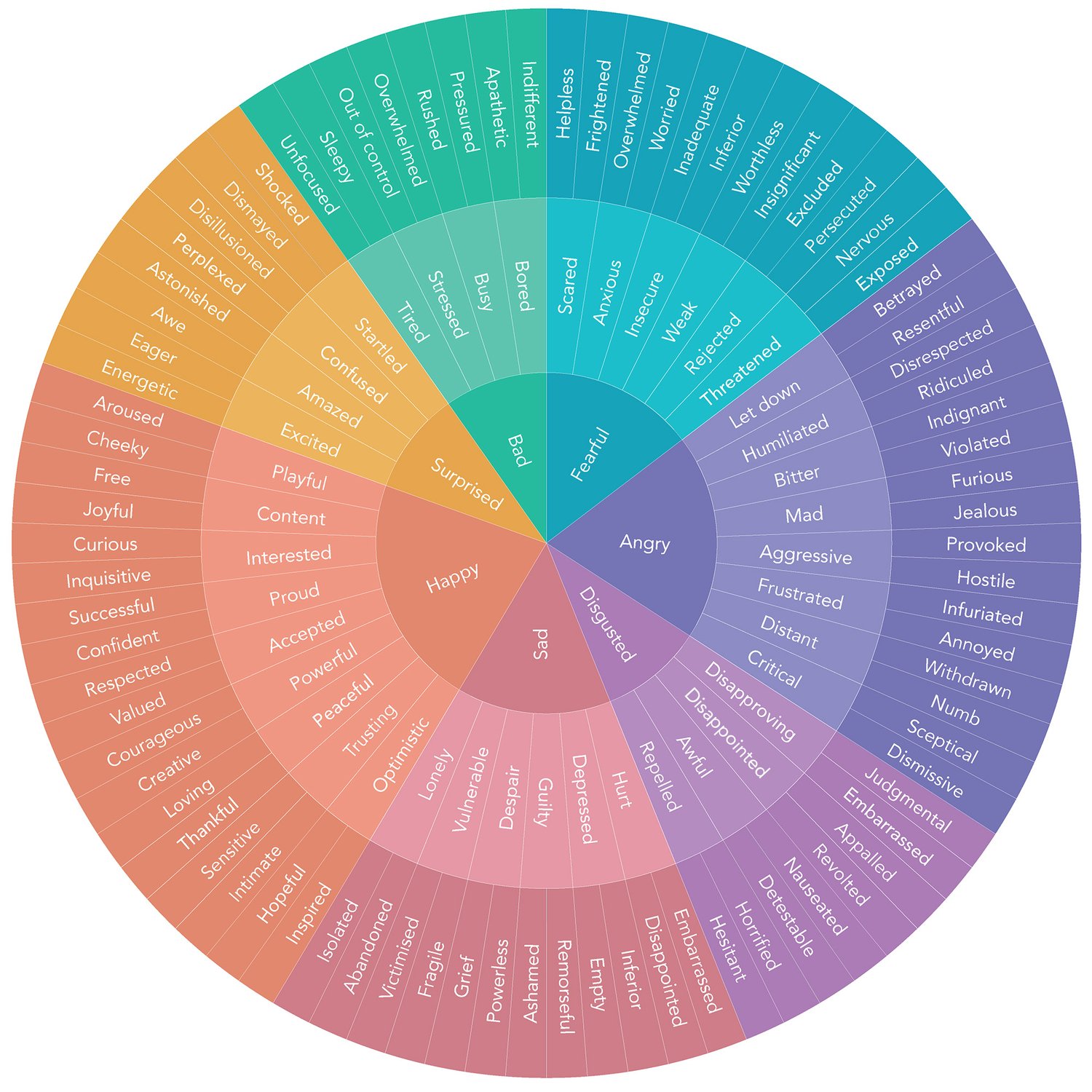}
		}
	\end{center}
	\caption{Example of two emotion wheels.}
	\label{Figure4}
\end{figure}

\subsection{Evaluation Metrics}
This track does not restrict the number or category of labels during prediction. Therefore, traditional metrics like accuracy are not suitable for MER-FG. For evaluation purposes, we follow the previous work \cite{lian2024open} and compute results in two stages.

\subsubsection{Grouping}
First, we apply a three-level grouping strategy to reduce the impact of synonyms:

\textbf{M1.}
    We normalize different forms of emotion words to their base form. For example, we map \emph{happier} and \emph{happiness} to \emph{happy}. This function is denoted as $F(\cdot)$.
	
\textbf{M2.}
    We map synonyms to a unified label. For example, we map \emph{happy} and \emph{joyful} to \emph{happy}. We call this function $G(\cdot)$.
	
\textbf{M3.}
    The emotion wheel provides natural hierarchical grouping, with basic emotions located in the innermost layer and more nuanced labels arranged in the outer layers \cite{plutchik1980general}. Figure \ref{Figure4} shows two emotion wheels. First, we group the labels by their levels from the innermost to the outermost as $L_{w}^1$, $L_{w}^2$, and $L_{w}^3$. Next, we define a mapping function $m_{w}^{i\rightarrow j}(\cdot)$ that transforms labels from level $L_{w}^i$ to their corresponding labels in level $L_{w}^j$. We then introduce two grouping functions, $W_1$ and $W_2$. For $W_1$, all labels are mapped to their corresponding labels in $L_{w}^1$:
    \begin{equation}
    W_1 =
    \begin{cases}
    y_w, & \text{if}\; y_w \in L_{w}^1 \\
    m_{w}^{2\rightarrow1}(y_w), & \text{if}\; y_w \in L_{w}^2 \\
    m_{w}^{2\rightarrow1}(m_{w}^{3\rightarrow2}(y_w)), & \text{if}\; y_w \in L_{w}^3 \\
    \end{cases}
\end{equation}
For $W_2$, all labels are mapped to corresponding labels in $L_{w}^2$:
\begin{equation}
    W_2 =
    \begin{cases}
    \text{select one label in}\; m_{w}^{1\rightarrow2}(y_w), & \text{if}\; y_w \in L_{w}^1 \\
    y_w, & \text{if}\; y_w \in L_{w}^2 \\
    m_{w}^{3\rightarrow2}(y_w), & \text{if}\; y_w \in L_{w}^3 \\
    \end{cases}
\end{equation}
The above grouping functions can be summarized as:
\begin{equation}
L_1(\cdot) = W_1{\left(G\left(F\left(\cdot\right)\right)\right)},
\end{equation}
\begin{equation}
L_2(\cdot) = W_2{\left(G\left(F\left(\cdot\right)\right)\right)}.
\end{equation}

In this paper, we employ five emotion wheels $\{w_i\}_{i=1}^5$, following the approach of previous works \cite{lian2024open}. This process mitigates the influence of the choice of emotion wheels, thereby yielding more reliable evaluation results. Consequently, the aforementioned grouping function is dependent on the specific emotion wheel used. Therefore, we denote the final grouping functions as $L_{w_i}^1(\cdot)$ and $L_{w_i}^2(\cdot)$, where the former focuses on coarse-grained grouping, while the latter emphasizes more fine-grained grouping.

\subsubsection{Metrics}
For each sample, the number of labels is variable. Let the dataset consist of $N$ samples. For sample $x_i$, the true labels are denoted as $\mathbf{Y}_i$ and the predicted labels are denoted as $\mathbf{\hat{Y}}_i$. For a grouping function $M \in\{L_{w_i}^1, L_{w_i}^2\}$, we define the following evaluation metrics:
\begin{equation}
\mbox{Precision}_{\mbox{s}}^{M}  = \frac{1}{N}\sum_{i=1}^{N}\frac{\left|M( \mathbf{Y}_i ) \cap M(\mathbf{\hat{Y}}_i)\right|}{\left|M(\mathbf{\hat{Y}}_i)\right|},
\end{equation}
\begin{equation}
\mbox{Recall}_{\mbox{s}}^{M} = \frac{1}{N}\sum_{i=1}^{N}\frac{\left|M( \mathbf{Y}_i ) \cap M(\mathbf{\hat{Y}}_i)\right|}{\left|M(\mathbf{{Y}}_i)\right|},
\end{equation}
\begin{equation}
\mbox{F}_{\mbox{s}}^{M}  = 2\times\frac{\mbox{Precision}_{\mbox{s}}^{M} \times\mbox{Recall}_{\mbox{s}}^{M} }{\mbox{Precision}_{\mbox{s}}^{M} +\mbox{Recall}_{\mbox{s}}^{M} }.
\end{equation}
Here, $\mbox{Precision}_{\mbox{s}}^{M}$ represents the proportion of correctly predicted labels; $\mbox{Recall}_{\mbox{s}}^{M}$ measures whether the prediction covers all ground truth labels; and $\mbox{F}_{\mbox{s}}^{M}$ is the harmonic mean of these two metrics, providing a more comprehensive evaluation. In this paper, we define the following scores $S_1$ and $S_2$ based on different grouping functions, and use their average results for the final ranking:
\begin{equation}
S_1 = \text{Avg}[\mbox{F}_{\mbox{s}}^{M} ], M(\cdot) \in L_{w_i}^1(\cdot),
\end{equation}
\begin{equation}
S_2 = \text{Avg}[\mbox{F}_{\mbox{s}}^{M} ], M(\cdot) \in L_{w_i}^2(\cdot).
\end{equation}

\subsection{Baseline Framework}
\subsubsection{Zero-shot Baselines}
The primary objective of MER-FG is to generate appropriate emotion labels for a given sample without being constrained by a predefined emotion taxonomy. Consequently, LLM-driven baselines are well-suited for this task, as they have extensive vocabularies that enable the generation of fine-grained emotion labels. Given that emotions are often expressed through multimodal cues, we primarily select multimodal LLMs (MLLMs) as our baselines, including representative frameworks such as SALMONN and Chat-UniVi. To extract emotion-related clues, we employ the following prompt: \emph{\textcolor[rgb]{0.93,0.0,0.47}{{As an expert in the field of emotions, please focus on the facial expressions, body movements, environment, acoustic information, subtitle content, etc., in the video to discern clues related to the emotions of the individual. Please provide a detailed description and ultimately predict the emotional state of the individual in the video.}}} 

Next, we use Qwen2.5 to extract emotion labels from the above descriptions using the following prompt: \emph{\textcolor[rgb]{0.93,0.0,0.47}{{Please assume the role of an expert in the field of emotions. We provide clues that may be related to the emotions of the characters. Based on the provided clues, please identify the emotional states of the main characters. Please separate different emotional categories with commas and output only the clearly identifiable emotional categories in a list format. If none are identified, please output an empty list.}}} 

Finally, we obtain the emotion prediction results for MER-FG. This process does not involve any training and operates in a zero-shot setup. For MLLMs, we use their 7B parameter weights by default. All models are implemented in PyTorch, and all inference processes are conducted on an A100 GPU. Additionally, we leverage the vLLM toolkit to accelerate the inference process.

\subsubsection{AffectGPT}
We also evaluate the performance of an emotion-specific MLLM, AffectGPT \cite{lian2025affectgpt}. This framework employs a pre-fusion mechanism to enhance multimodal integration. During our experiments, we train the model on two datasets: OV-MERD and MER-Caption+. All pretrained models are available in our official baseline code repository. For detailed implementation instructions, please refer to the baseline code.

\subsection{Baseline Results}

Table \ref{Table4} presents all baseline results. In Table \ref{Table4}, AffectGPT outperforms the zero-shot baselines, achieving higher scores in both coarse-grained scores $S_1$ and fine-grained scores $S_2$. Additionally, the model trained on MER-Caption+ demonstrates superior performance compared to the one trained on OV-MERD. These findings indicate that although OV-MERD has high-quality labels, its limited dataset size is insufficient to support effective training. In contrast, MER-Caption+ offers a larger number of samples with relatively accurate labels, resulting in better performance on MER-FG.

\begin{table}[t]
	\centering
	\caption{MER-FG baseline results.}
	\label{Table4}
    \scalebox{0.9}{
	\begin{tabular}{l|cc>{\columncolor[gray]{0.9}}c}
		\hline
		\multirow{2}{*}{\textbf{Model}} & \multicolumn{3}{c}{\textbf{Metrics}} \\
		& $S_1$ $(\uparrow)$ & $S_2$ $(\uparrow)$ & Avg $(\uparrow)$ \\
		\hline
Otter                   & 14.64 & 04.46 & 09.55 \\
Video-LLaVA             & 27.40 & 12.18 & 19.79 \\
Qwen-Audio              & 28.22 & 16.27 & 22.25 \\
VideoChat2              & 34.07 & 17.78 & 25.92 \\
Video-ChatGPT           & 35.29 & 19.77 & 27.53 \\
LLaMA-VID               & 40.89 & 21.60 & 31.25 \\
SALMONN                 & 41.33 & 22.50 & 31.92 \\
Chat-UniVi              & 43.33 & 23.90 & 33.62 \\
VideoChat               & 43.48 & 24.30 & 33.89 \\
mPLUG-Owl               & 46.28 & 27.32 & 36.80 \\
\hline
AffectGPT(OV-MERD) \cite{lian2025affectgpt}      & 47.81 & 30.16 & 38.98 \\
AffectGPT(MER-Caption+) \cite{lian2025affectgpt}  & \textbf{57.36} & \textbf{36.35} &\textbf{46.86} \\
		\hline
	\end{tabular}
    }
\end{table}

\section{MER-DES}

\subsection{Dataset}
\begin{figure}[t]
	\begin{center}
		\subfigure[OV-MERD]{
			\label{Figure5-1}
			\centering
			\includegraphics[width=0.476\linewidth, trim=0 0 0 0]{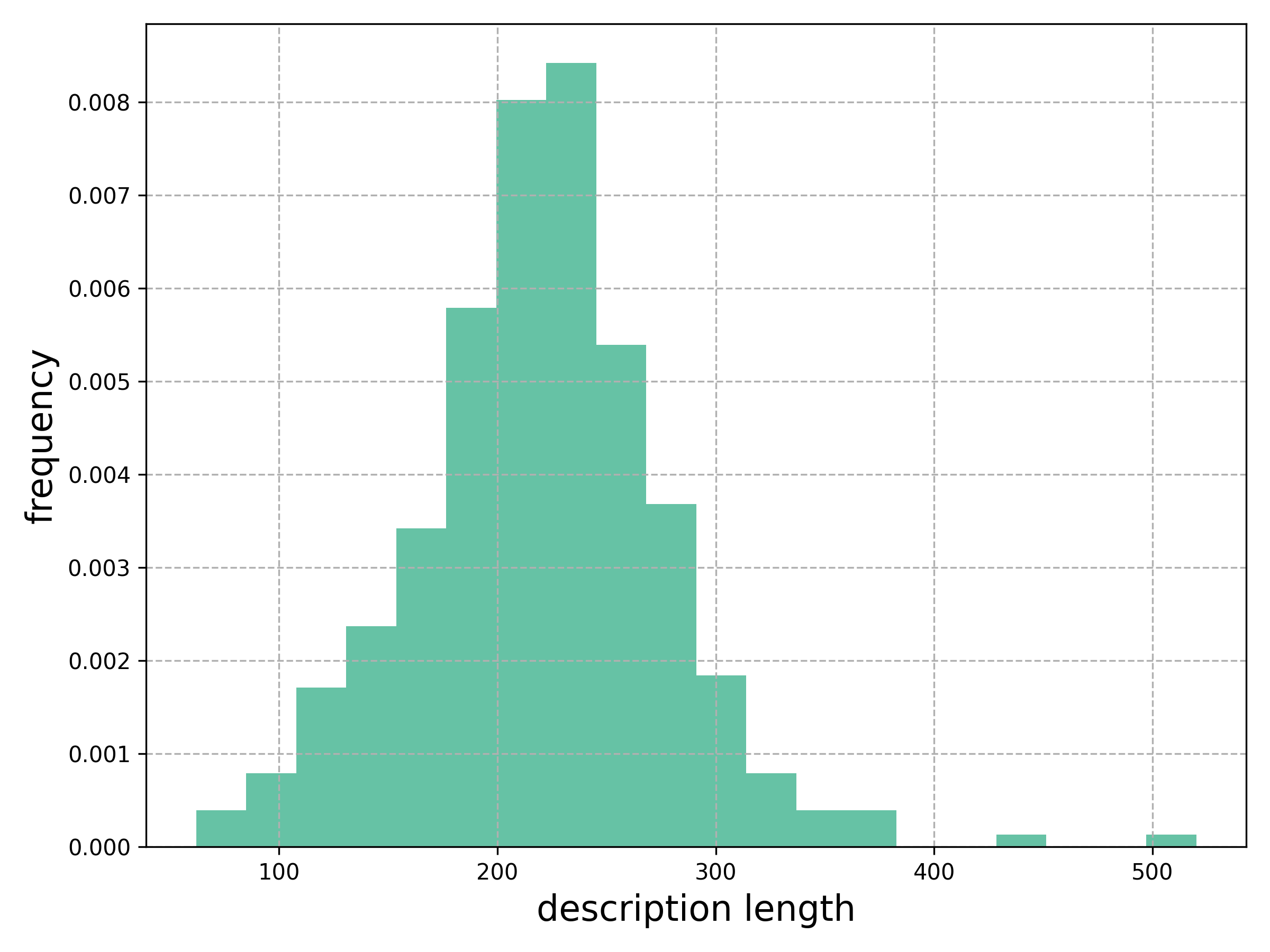}
		}
		\subfigure[MER-Caption+]{
			\label{Figure5-2}
			\centering
			\includegraphics[width=0.476\linewidth, trim=0 0 0 0]{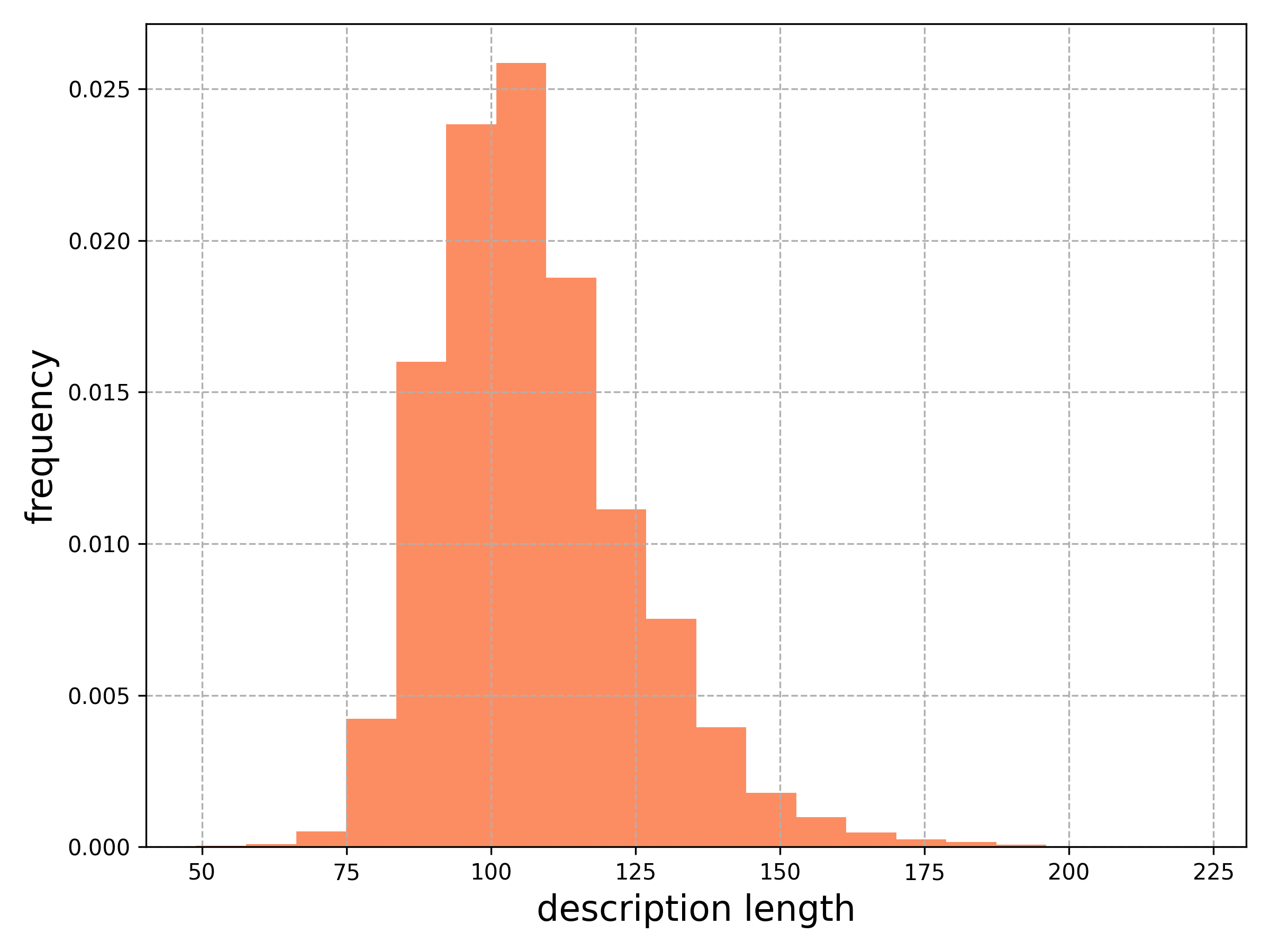}
		}
	\end{center}
	\caption{Emotion description length distribution.}
	\label{Figure5}
\end{figure}

ER-SEMI and MER-FG focus on emotion word prediction, whereas MER-DES extends beyond emotion words by integrating multimodal clues to enhance the interpretability of each emotion label. Figure \ref{Figure1} illustrates the task differences among these tracks. Since OV-MERD and MER-Caption+, which are used in MER-FG, also provide emotion descriptions, they are employed as the training set for MER-DES. Figure \ref{Figure5} presents the distribution of description lengths in these datasets. We observe that both datasets provide detailed emotional descriptions for each sample.

\subsection{Evaluation Metrics} 
MER-DES leverages multimodal cues to enhance the interpretability of each emotion label. Given that interpretability is inherently a relatively subjective metric, we plan to recruit annotators to manually evaluate the quality of the emotion descriptions. Simultaneously, we will invite two members from each team to assist with the scoring process. The evaluation will be conducted along two dimensions:

First, we will evaluate whether the submission results include interpretable clues. This criterion distinguishes MER-DES from MER-SEMI and MER-FG by ensuring that participants do not merely provide emotion labels without supporting evidence.

Second, we will assess the quality of the emotion descriptions from two perspectives: 1) whether the provided emotion clues are present in the video; and 2) whether the emotion and its associated clues are logically connected.

Initially, we plan to provide real descriptions and calculate similarity scores between real and predicted descriptions, as done in EMER \cite{lian2023explainable}. However, human emotional expression is complex and difficult to fully capture all emotion-related visual and acoustic cues. Therefore, we shift our focus to manual evaluation. During the ranking process, each team is required to participate in the labeling process. We will exclude annotators with low inter-annotator agreement compared to others to ensure annotation quality. Additionally, we will mask the sample names and require participants to submit results in English while using their translated Chinese versions for ranking. This approach prevents participants from identifying their own submissions and deliberately assigning inflated scores. The final rankings will be determined based on these manual annotations.

\subsection{Baseline Framework}
\subsubsection{Zero-shot Baselines}
You can also use MLLMs as baselines. Specifically, you can prompt the MLLMs with the following instructions: \emph{\textcolor[rgb]{0.93,0.0,0.47}{{As an expert in the field of emotions, please focus on the facial expressions, body movements, environment, acoustic information, subtitle content, etc., in the video to discern clues related to the emotions of the individual. Please provide a detailed description and ultimately predict the emotional state of the individual in the video.}}}

\subsubsection{AffectGPT}
You can also fine-tune AffectGPT using the emotion descriptions from the OV-MERD and MER-Caption+ datasets. \textbf{The official GitHub repository already includes baseline code and pretrained weights for this track.}

\begin{table}[t]
	\centering
	\caption{Dataset statistics for MER-PR.}
	\label{Table5}
    \scalebox{0.9}{
	\begin{tabular}{l|ccc|c}
		\hline
		& Train & Val & Test & Total \\
		\hline 
		Subjects & 153 & 40 & 40 & 233\\
            \# Samples & 2,448 & 640 & 640 & 3,728 \\
		\hline
	\end{tabular}
    }
\end{table}

\begin{table*}[t]
	\centering
	\caption{Unimodal results for MER-PR. In this table, the abbreviations ``O'', ``C'', ``E'', ``A'', and ``N'' represent the \emph{openness}, \emph{conscientiousness}, \emph{extraversion},  \emph{agreeableness}, and \emph{neuroticism}, respectively.}
	\label{Table6}
\scalebox{0.9}{
	\begin{tabular}{c|cccccc|ccccc>{\columncolor[gray]{0.9}}c}
\hline
\multirow{2}{*}{Feature} & \multicolumn{6}{c|}{Val} & \multicolumn{6}{c}{Test} \\
\cline{2-13} 
& O$(\downarrow)$ & C $(\downarrow)$ & E $(\downarrow)$ & A $(\downarrow)$ & N $(\downarrow)$ & Avg $(\downarrow)$ & O $(\downarrow)$& C $(\downarrow)$ & E $(\downarrow)$& A $(\downarrow)$ & N$(\downarrow)$ & Avg $(\downarrow)$\\ 
\hline
VIT             & 0.179            & \textbf{0.159}  & 0.174           & 0.231           & 0.183           & 0.185            & \textbf{0.145}   & 0.106           & 0.178           & 0.217           & \textbf{0.156} & 0.160\\
ClipVIT-B16     & 0.171            & 0.169           & 0.177           & 0.225           & 0.187           & 0.185            & 0.158            & 0.108           & 0.180           & 0.220           & 0.165          & 0.166\\
ClipVIT-L14     & 0.172            & 0.168           & 0.180           & 0.230           & \textbf{0.180}  & 0.186            & 0.173            & 0.114           & 0.176           & 0.229           & 0.166          & 0.172\\ \hline
HUBERT-base     & 0.173            & 0.162           & 0.176           & 0.228           & 0.184           & 0.185            & 0.160            & \textbf{0.098}  & 0.183           & 0.218           & 0.167          & 0.165\\
HUBERT-large    & 0.200            & 0.171           & 0.194           & 0.240           & 0.200           & 0.201            & 0.189            & 0.150           & 0.210           & 0.228           & 0.174          & 0.190\\
Wav2vec2-base   & 0.169            & 0.160           & \textbf{0.172}  & \textbf{0.221}  & 0.182           & \textbf{0.180}   & 0.159            & 0.092           & 0.177           & 0.215           & 0.158          & 0.160\\
Wav2vec2-large  & 0.198            & 0.168           & 0.190           & 0.237           & 0.200           & 0.198            & 0.190            & 0.146           & 0.205           & 0.223           & 0.173          & 0.187\\
WavLM-base      & 0.178            & 0.164           & 0.179           & 0.227           & 0.184           & 0.186            & 0.156            & 0.101           & 0.180           & 0.219           & 0.162          & 0.164\\
WavLM-large     & 0.191            & 0.174           & 0.191           & 0.246           & 0.206           & 0.201            & 0.200            & 0.147           & 0.209           & 0.251           & 0.205          & 0.202\\ \hline
Sentence-BERT   & 0.181            & 0.181           & 0.176           & 0.234           & 0.181           & 0.190            & 0.159            & 0.104           & 0.181           & 0.219           & 0.163          & 0.165\\
ChatGLM2-6B     & 0.188            & 0.180           & 0.195           & 0.248           & 0.202           & 0.203            & 0.182            & 0.133           & 0.200           & 0.240           & 0.180          & 0.187\\
Baichuan-13B    & \textbf{0.167}   & 0.160           & \textbf{0.172}  & 0.229           & \textbf{0.180}  & 0.182            & 0.156            & \textbf{0.098}  & \textbf{0.157}  & \textbf{0.214}  & 0.157          & \textbf{0.156} \\ 
\hline
\end{tabular}
}
\end{table*}

\begin{table*}[t]
	\centering
	\caption{Multimodal results for MER-PR.}
	\label{Table7}
\scalebox{0.9}{
	\begin{tabular}{ccc|cccccc|ccccc>{\columncolor[gray]{0.9}}c}
\hline
\multicolumn{3}{c|}{Features} & \multicolumn{6}{c|}{Val} & \multicolumn{6}{c}{Test} \\
\cline{4-15}
V & A & T & O$(\downarrow)$ & C $(\downarrow)$ & E $(\downarrow)$ & A $(\downarrow)$ & N $(\downarrow)$ & Avg $(\downarrow)$ & O $(\downarrow)$& C $(\downarrow)$ & E $(\downarrow)$& A $(\downarrow)$ & N$(\downarrow)$ & Avg $(\downarrow)$\\ 
\hline
VIT & W2V & --- & 0.171          & \textbf{0.160}  & 0.175          & \textbf{0.223}   & 0.185          & 0.183          & 0.159           & 0.095            & 0.177          & 0.213           & \textbf{0.158}   & 0.160\\
--- & W2V & BAI & 0.170          & \textbf{0.160}  & \textbf{0.171} & \textbf{0.223}   & 0.184          & \textbf{0.182} & 0.159           & 0.096            & 0.175          & 0.215           & 0.160            & 0.161\\
VIT &---  & BAI & \textbf{0.167} & 0.161           & 0.172          & 0.228            & \textbf{0.182} & \textbf{0.182} & \textbf{0.154}  & 0.097            & 0.175          & 0.216           & \textbf{0.158}   & 0.160\\
VIT & W2V & BAI & 0.172          & \textbf{0.160}  & 0.175          & 0.224            & 0.187          & 0.184          & 0.156           & \textbf{0.093}   & \textbf{0.171} & \textbf{0.212}  & 0.160            & \textbf{0.158} \\
\hline
\end{tabular}
}
\end{table*}

\section{MER-PR}

\subsection{Dataset}
This track aims to explore whether emotion prediction results can enhance the performance of personality recognition. For this purpose, we utilize the MDPE dataset \cite{cai2024mdpe}, which provides annotations for both emotion and personality traits. The dataset and baseline code are available on GitHub\footnote{\emph{https://github.com/cai-cong/MER25\_personality}}. Table \ref{Table5} provides dataset statistics. 

For personality traits, we used a Big Five personality questionnaire consisting of 60 items. Responses were used to derive scores for five personality dimensions (\emph{openness}, \emph{conscientiousness}, \emph{extraversion},  \emph{agreeableness}, and \emph{neuroticism}), with each dimension represented as a floating-point value ranging from 0 to 1. The dataset includes 233 participants, all native Chinese speakers from diverse backgrounds.

For emotion labels, each participant watched 16 emotion-induction videos, comprising two videos designed to elicit each of the eight target emotions: \emph{sadness}, \emph{happiness}, \emph{relaxation}, \emph{surprise}, \emph{fear}, \emph{disgust}, \emph{anger}, and \emph{neutral}. After watching, participants described their emotional responses and completed a self-report emotion scale to quantify the intensity of each emotion on a 1–5 Likert scale (1 = no emotion, 5 = strongest emotion). This process collected 16 samples per participant, resulting in a total of 3,728 samples.

\subsection{Evaluation Metric}
This track focuses on personality recognition. In the dataset, personality traits are quantified using five floating-point values. For evaluation, we employ the Root Mean Square Error (RMSE) metric to measure the discrepancy between true and predicted personality scores. The RMSE is calculated as follows:
\begin{equation}
    \text{RMSE} = \sqrt{\frac{1}{N} \sum_{i = 1}^{N} \left( y_{i} - \hat{y}_{i} \right)^{2}},
\end{equation}
where $N$ is the dataset size. $y_{i}$ and $\hat{y}_{i}$ denote the true and predicted personality scores, respectively.

\subsection{Baseline Framework}
For unimodal features, we employ fully connected layers to extract hidden representations and predict personality scores:
\begin{equation}
h_{i}^{m}=\text{ReLU}\left( f_{i}^ {m}W_ {m}^{h}+b_{m}^{h} \right) ,{m\in \left\{ a,l,v \right\}},
\end{equation}
\begin{equation}
\hat {y}_i  =  h_{i}^ {m}W_ {m}^{d}+b_{m}^{d}, m\in \left\{ a,l,v \right\},
\end{equation}
where $h_{i}^{m} \in \mathbb{R}^h $ is the hidden feature for each modality, and $\hat {y}_i \in \mathbb{R}^5 $ is the estimated personality probabilities. 

For multimodal features, we use concatenation for feature fusion:
\begin{equation}
h_i= \text{Concat}\left( h_{i}^{a},h_{i}^{l},h_{i}^{v} \right),
\end{equation}
\begin{equation}
\alpha_i=  h_{i}^ {T}W_ {\alpha}+b_{\alpha},
\end{equation}
where $\alpha_i \in \mathbb{R}^5 $ is the estimated personality probabilities.

\subsection{Baseline Results}

\subsubsection{Unimodal Results}
We establish a unimodal benchmark for personality trait prediction across visual, acoustic, and textual modalities (see Table \ref{Table6}). For the visual modality, the ViT feature achieves competitive performance, particularly in predicting \emph{conscientiousness}. Among acoustic features, Wav2Vec2-base yields the highest average results. The textual modality outperforms all other unimodal features, with Baichuan-13B achieving the best average scores on the test set (0.156), excelling in \emph{extraversion} (0.157), \emph{agreeableness} (0.214), and \emph{neuroticism} (0.157). Meanwhile, although textual cues are the most informative, visual and acoustic signals provide complementary trait-specific insights. For instance, ViT is particularly effective for \emph{conscientiousness}, whereas Wav2Vec2-base performs well for \emph{agreeableness}.

\subsubsection{Multimodal Results} 
In Table \ref{Table7}, we present the multimodal fusion results based on several best-performing unimodal features. While multimodal fusion does not consistently outperform individual modalities in terms of average test scores, it yields improvements for specific personality traits. This underscores the potential advantages of tailoring fusion strategies to individual traits.

\section{Challenge Protocol}
For \textbf{MER-SEMI}, \textbf{MER-FG}, and \textbf{MER-DES}, to download the dataset, participants must complete an End User License Agreement (EULA) available on Hugging Face\footnote{\emph{https://huggingface.co/datasets/MERChallenge/MER2025}}. The EULA clearly states that the dataset is for academic research purposes only and prohibits any modifications or uploads to the Internet. We emphasize that manual annotation of MER2025 samples is strictly prohibited. For these tracks, the test samples are selected from the 124k unlabeled samples (see Table \ref{Table1}). To reduce the task difficulty, we reduce the evaluation scope from 124k to 20k samples. Participants must submit predictions for these 20k samples, which cover all the test samples of these tracks. As shown in Figure \ref{Figure1}, each track has distinct objectives. For MER-SEMI, participants must predict the most likely label from six predefined categories: \emph{worried}, \emph{happy}, \emph{neutral}, \emph{angry}, \emph{surprised}, and \emph{sad}; For MER-FG, participants can freely predict any emotion labels without restrictions on category or quantity; For MER-DES, participants are required to submit both multimodal evidence and corresponding emotion labels to improve model interpretability.

For \textbf{MER-PR}, the dataset and baseline code are available on GitHub\footnote{\emph{https://github.com/cai-cong/MER25\_personality}}. We provide an official test set, and participants can directly submit predictions for the test set.

\section{Conclusions}
This year's MER2025 focuses on the theme ``When Affective Computing Meets Large Language Models'' and contains four distinct tracks. This paper presents the datasets, baselines, evaluation metrics, and experimental results for all tracks. For MER-SEMI, we evaluate various unimodal features and multimodal fusion approaches, providing a strong baseline for categorical emotion recognition under a fixed taxonomy. For MER-FG and MER-DES, we employ LLM-driven baselines to generate fine-grained emotions with corresponding multimodal clues, ensuring accurate and interpretable emotion understanding. For MER-PR, we test diverse unimodal features for personality detection. To promote reproducibility, we have open-sourced the code and pretrained models on our official GitHub repository. We hope all participants enjoy this year’s challenge! Your continued support and engagement make this challenge truly meaningful.

%%
%% The next two lines define the bibliography style to be used, and
%% the bibliography file.
\bibliographystyle{ACM-Reference-Format}
\bibliography{mybib}

\end{document}